\documentclass{article}

\usepackage[utf8]{inputenc} 
\usepackage[T1]{fontenc}    
\usepackage{hyperref}       
\usepackage{url}            
\usepackage{booktabs}       
\usepackage{amsfonts}       
\usepackage{nicefrac}       
\usepackage{microtype}      
\usepackage[acronym]{glossaries}    
\usepackage{glossaries}     
\usepackage{graphicx}
\usepackage{subfig}
\usepackage{amsmath}
\newcommand{\norm}[1]{\left\lVert#1\right\rVert} 
\usepackage{float}

\author{
  Ricky Chen\thanks{Equal contribution.}\\
  Simon Fraser University \\
  \texttt{shuo\_chen\_4@sfu.ca}
  \and
  Timothy T. Yu$^{\ast}$\\
  Simon Fraser University \\
  \texttt{timothy\_yu@sfu.ca}
  \and
  Gavin Xu$^{\ast}$\\
  Simon Fraser University \\
  \texttt{jiacheng\_xu\_2@sfu.ca}
  \and
  Da Ma\\
  Simon Fraser University \\
  \texttt{da\_ma@sfu.ca}
  \and
  Marinko V. Sarunic\\
  Simon Fraser University \\
  \texttt{msarunic@sfu.ca}
  \and
  Mirza Faisal Beg\\
  Simon Fraser University \\
  \texttt{faisal\_beg@sfu.ca}
}

\title{Domain Adaptation via CycleGAN for Retina Segmentation in Optical Coherence Tomography}

\newglossaryentry{bscan}
{
        name=B-scan,
        description={cross sectional slice by scanning through the volume in a raster scan pattern}
}
\newacronym{ai}{AI}{Artificial Intelligence}
\newacronym{cyclegan}{CycleGAN}{Cycle-Consistent Generative Adversarial Networks}
\newacronym{fcn}{FCN}{Fully Convolutional Network}
\newacronym{irf}{IRF}{Intraretinal Fluid}
\newacronym{oct}{OCT}{Optical Coherence Tomography}
\newacronym{ped}{PED}{Pigment Epithelium Detachments}
\newacronym{relu}{ReLU}{Rectified Linear Unit}
\newacronym{srf}{SRF}{Subretinal Fluid}
\newacronym{ukb}{UKB}{UK Biobank dataset}
\newacronym{ilm}{ILM}{inner limiting membrane}
\newacronym{rpe}{RPE}{retinal pigment epithelium}
\newacronym{dsc}{DSC}{Dice Similarity Coefficient}
\newacronym{js}{JS}{Jaccard Score}
\newacronym{auc}{AUC}{area under the curve of the receiver operating characteristic}

\begin{document}

\maketitle

\begin{abstract}
  With the FDA approval of \acrfull{ai} for point-of-care clinical diagnoses \cite{ADA}\cite{abramoff}, the generalizability of a network’s performance is of the utmost importance as clinical decision-making must be domain-agnostic \cite{generalizability}. A method of tackling the problem is to increase the dataset to include images from a multitude of domains; while this technique is ideal, the security requirements of medical data is a major limitation. Additionally, researchers with developed tools benefit from the addition of open-sourced data, but are limited by the difference in domains. Herewith, we investigated the implementation of a \acrfull{cyclegan} for the domain adaptation of \acrfull{oct} volumes. This study was done in collaboration with the Biomedical Optics Research Group and Functional \& Anatomical Imaging \& Shape Analysis Lab at Simon Fraser University. In this study, we investigated a learning-based approach of adapting the domain of a publicly available dataset, \acrfull{ukb} \cite{ADA}. To evaluate the performance of domain adapatation, we utilized pre-existing retinal layer segmentation tools developed on a different set of RETOUCH \acrshort{oct} data. This study provides insight on state-of-the-art tools for domain adaptation compared to traditional processing techniques as well as a pipeline for adapting publicly available retinal data to the domains previously used by our collaborators. 

\end{abstract}

\section{Introduction}
\label{l_intro}
\acrfull{ai} has caused a shift in focus towards the role of data in decision-making, and is projected to have an impact in the healthcare system. While \acrshort{ai} has been shown to improve accuracy, specificity, and sensitivity of clinician diagnoses \cite{BMJ}, there are many valid concerns and limitations of these systems. Alongside patient security \cite{ethics}, quality of care \cite{ethics}, and equality \cite{ethics}, the generalizability of these tools may be lacking \cite{generalizability}. As data ranges across different operators, acquisition systems, and patients, the domain-dependent performance of the deep learning tools vary and must be validated in the environment that it is used \cite{generalizability}. In this study, we investigate the use of a \acrfull{cyclegan} to adapt data from an external domain to the domain that our learning-based tools have been trained on. The performance of domain adaptation from \acrfull{ukb} to RETOUCH was evaluated through retina segmentation from the \acrfull{ilm} to the \acrfull{rpe}. 

\subsection{Data Acquisition}
Two datasets were used for domain adaptation. The RETOUCH dataset, which was originally published in a half day challenge, contains retinal \acrfull{oct} volumes along with labels of the retinal layers, \acrfull{irf}, \acrfull{srf}, and \acrfull{ped} \cite{RETOUCH-dataset}. The volumes were imaged using a Topcon machine, and each volume consists of 128 \gls{bscan}s of size 650x512. This study utilized 13 volumes in total, which is equivalent to 1664 \gls{bscan}s. \acrlong{ukb} was released by \acrshort{ukb}, which contains in-depth genetic and health information from half a million UK participants \cite{UKB-dataset}. More than 80000 volumes were available in the \acrshort{ukb} dataset, but only 13 volumes without pathological cases were chosen to be consistent with the RETOUCH dataset. Figure 1 shows sample \gls{bscan}s from the two domains.

\begin{figure}[h]
  \centering
  \subfloat[RETOUCH sample B-scan]{\includegraphics[scale=0.3]{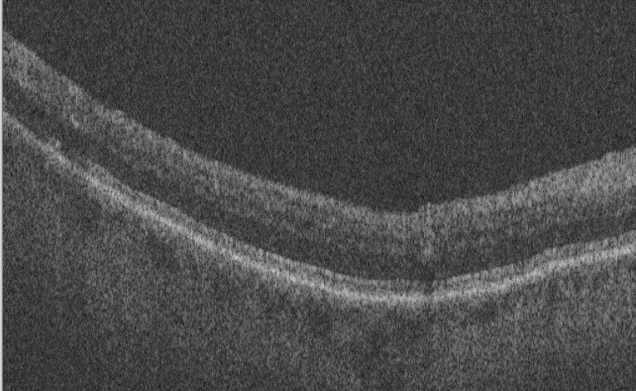}}
  \hspace{0.3cm}
  \subfloat[UKB sample B-scan]{\includegraphics[scale=0.3]{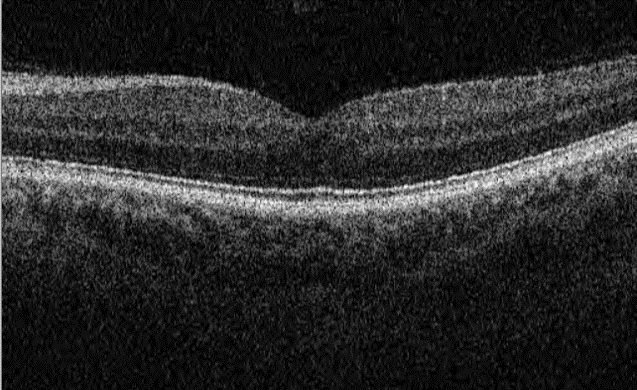}}
  \caption{Comparison of two datasets/domains}
\end{figure}

As shown in the above figure, \acrshort{ukb} appears to have undergone some de-noising processes, which may include but is not limited to: 2D retina flattening, pepper noise reduction, etc. 

\subsection{Layer Segmentation Network}
The LF-Unet, a deep neural network based on U-Net and fully convolutional network, was used as a pre-trained layer segmentation network \cite{LF-UNet}. The architecture of the U-Net leverages a symmetrical convolutional network. It consists of multiple 3x3 convolutional layers followed by a \acrfull{relu} and 2x2 max pooling layer \cite{UNet}. The up-sampled output is then concatenated with the corresponding cropped feature maps. The network can thus learn multiple in-depth features and perform concrete pixel-wise segmentation. In addition to a simple U-Net, LF-UNet cascades a \acrfull{fcn}. The \acrshort{fcn} is fed the features from both contracting block and expansive block, which has been shown to perform better in boundary detection \cite{UNet}. The LF-UNet was trained on the RETOUCH dataset with the best state-of-art evaluation scores reported by MICCAI 2017 Satellite Event \cite{RETOUCH-dataset}.

\section{Methodologies}
\label{l_method}

\subsection{Traditional Method}
\begin{figure}[ht]
\centering
\includegraphics[scale=0.4]{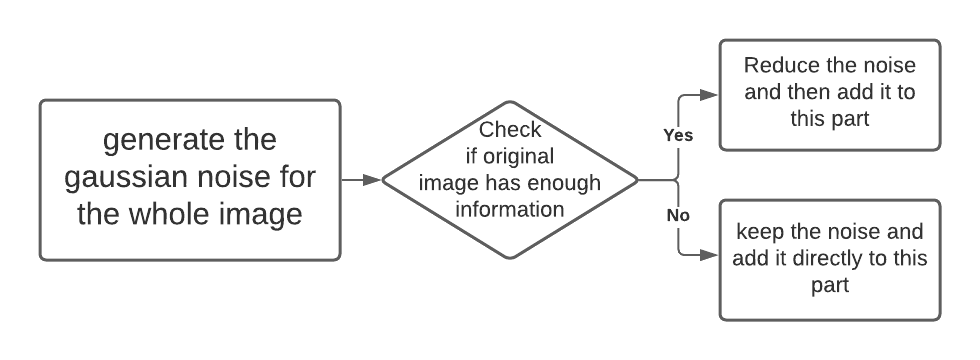}
\caption{Traditional Mathematical Method}
\label{fig:mathemathicalMethod}
\end{figure}

Through visual inspection, the noise and intensity differences are evident between the two domains. Thus, without using machine learning, we inject Gaussian noise to visually match the two domains of data.
The simple formula of Gaussian noise is:
\[P(x) = \frac{1}{{\sigma \sqrt {2\pi } }}e^{{{ - \left( {x - \mu } \right)^2 } \mathord{\left/ {\vphantom {{ - \left( {x - \mu } \right)^2 } {2\sigma ^2 }}} \right. \kern-\nulldelimiterspace} {2\sigma ^2 }}}\]
We determined which regions to inject Gaussian noise by utilizing a simple density detection technique. If a pixel's density was larger than a specified threshold, we set the intensity to 196 (from 0 to 255) and no noise was added. Furthermore, if the value was larger than 225 and the neighbouring intensity density was low, noise was not added to the adjacent pixels.
After determining the regions to inject noise, we introduced Gaussian noise accordingly to generate our traditionally processed images.
The logic of this method is detailed in Figure \ref{fig:mathemathicalMethod}.

\subsection{CycleGAN}
\acrshort{cyclegan} was initially introduced in 2017, to learn bi-directional mapping functions between two different domains \cite{Cycle-GAN}. It emphasizes the concept of cycle consistency, where the reconstructed image obtained by a cyclic adaptation is expected to be identical to the original image. This regularization-purpose approach allows the adaptation for both domains, and it is widely used in many data-related applications.

\subsubsection{Architecture}
The architecture and main components of the network are shown below:

\begin{figure}[ht]
\centering
\includegraphics[width=0.4\textwidth,height=5cm]{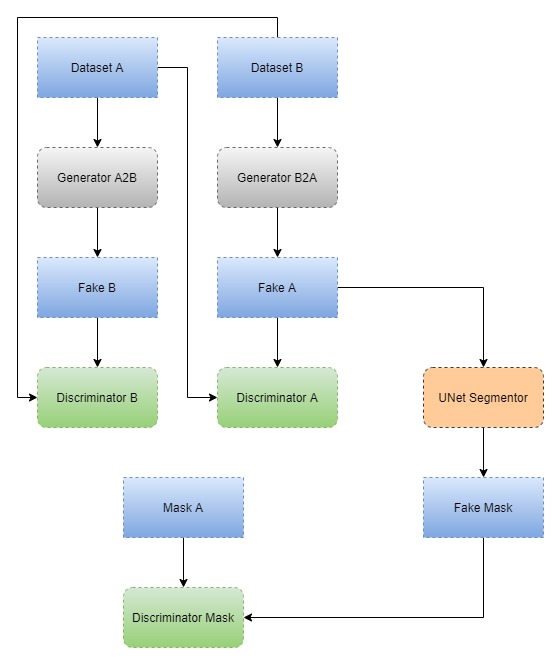}
\caption{Flow diagram}
\label{fig:flow}
\end{figure}

As shown in Figure \ref{fig:flow}, \acrshort{cyclegan} consists of two main deep neural network blocks. Specifically, the generator and discriminator. The generator adapts the data from one domain to another, while the discriminator attempts to distinguish the generated images. DatasetA refers to RETOUCH B-scans, while datasetB refers to UKB B-scans. DatasetA was passed into the generatorA2B, then the generated images were passed to the discriminatorB along with the real datasetB. The same data-flow existed for datasetB, generatorB2A and discriminatorA, respectively. In addition to the basic \acrshort{cyclegan}, the segmentation data-flow was included. Since the LF-UNet was trained for segmenting B-scans from datasetA, the segmentation generated from both datasetA and generated image A was passed into the discriminator. The losses from both generators and discriminators were updated every epoch.

\begin{figure}[H]
  \centering
  \subfloat[Generator]{\includegraphics[width=0.3\textwidth,height=8cm]{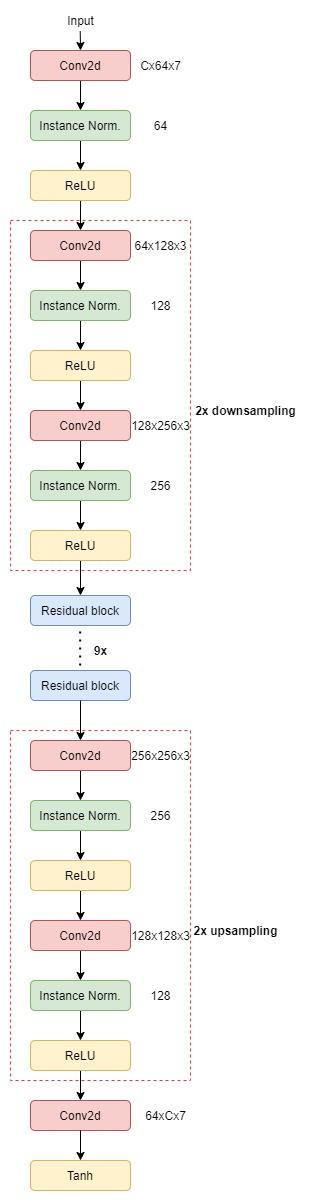}}
  \hfill
  \subfloat[Discriminator]{\includegraphics[width=0.3\textwidth,height=8cm]{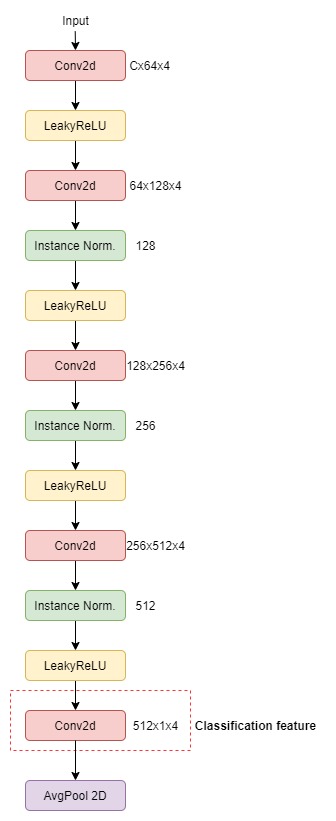}}
  \hfill
  \subfloat[Residual block]{\includegraphics[width=0.2\textwidth,height=6cm]{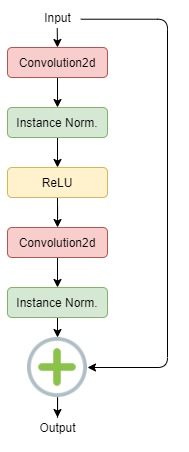}}
  \caption{CycleGAN architecture and components}
  \label{fig:cyclegan}
\end{figure}

As shown in Figure \ref{fig:cyclegan} (a)(b), a detailed architecture for the core blocks is illustrated. The generator began with a 2x down-sampling operation for extracting deep features, then the feature maps were passed into 9 consecutive residual blocks, which is shown in Figure \ref{fig:cyclegan} (c). The residual block performs two sets of convolution and normalization layers with a \acrshort{relu} activation function in between. The output was element-wise added with the input via a skip-connection to allow the network to learn simple functions like identity function that may be difficult to capture using convolutional filters. After the series of residual blocks, the 2-x up-sampling operations were performed to reconstruct the feature maps in the other domain. The hyperbolic tangent activation (tanh) function was applied as the final rectifier. The discriminator was simpler than the generator and conducted 4-level feature reconstructions, followed by a 2D average pooling layer. 

\subsubsection{Loss functions}
A Pytorch-implemented version was adopted as the baseline loss function for training the two OCT datasets \cite{Pytorch-CycleGAN}. The original paper mentioned two loss functions. Minimax loss, or adversarial loss, is designed to maximize the error of the discriminators. Given both generator G and discriminator D, and mapping function G: X->Y, we have:
\[L_{GAN}(G,D_Y,X,Y) = E_{y\sim p_{data}(y)}[logD_Y(y)] + E_{x\sim p_{data}(x)}[log(1-D_Y(G(x))]\]
Also, for the reverse mapping function F: Y->X, we have another one: 
\[L_{GAN}(F,D_X,Y,X) = E_{x\sim p_{data}(x)}[logD_X(x)] + E_{y\sim p_{data}(y)}[log(1-D_X(F(y))]\]
The loss is minimized when the discriminator successfully distinguishes the input image. Therefore, a higher discriminator loss indicates a better performing generator. Cycle consistency loss is defined based on the image reconstruction on both forward and backward direction:
\[L_{cyc}(G,F) = E_{x\sim p_{data}(x)}[\norm{F(G(x))-x}_1] + E_{y\sim p_{data}(y)}[\norm{G(F(y))-y}_1]\]
As the cycle loss decreases, the adapted image becomes more similar to the original image. 

Two additional loss functions were added for better regularization. The identity loss was designed such that the generator performed no adaptation if the given input was from the output domain:
\[L_{id}(G,F) = E_{x\sim p_{data}(x)}[\norm{F(x)-x}_1] + E_{y\sim p_{data}(y)}[\norm{G(y)-y}_1]\]
The segmentation loss was designed to optimize such that the LF-UNet layer segmentation performance would be like that of the  segmentation using the ground truth, \acrshort{ukb}, data. The error was evaluated using both the dice loss and cross-entropy loss. Given the LF-UNet S and ground-truth segmentation M, we have:
\small\[L_{seg}(S,M,G) = E_{x\sim p_{data}(x)}[Dice(S(G(x))-S(x)] + E_{y\sim p_{data}(y)}[CE(S(G(x))-S(x))]\]
where the dice loss and cross-entropy loss is defined as:
\[Dice(A, B)=1 - \frac{2|A \cap B|}{|A|+|B|} \text{ and } CE(p, q)=-\sum_{\forall x} p(x) \log (q(x))\]
The dice loss ensured that the overlap area is maximized, while the cross-entropy loss handled the over-segmentation issue by maximizing the pixel-wise probabilities. 

\subsubsection{Fine-tuning}
The following fine-tuning process was inspired by multiple weighted loss approaches \cite{Better-cycles}. The author suggested three levels of regularization. First, the baseline network performs adaptation on pixel-wise aspect, where the feature maps may also guide the learning of the network. As shown in Figure \ref{fig:cyclegan} (b), the feature from the classification layer of the discriminator was extracted, which was added into the cycle consistency loss. Meanwhile, the feature weight $\gamma$ was applied to focus the network on the pixel-wise information during early stages of training as the early features from discriminator would not have good quality, but the feature-level information would be weighted higher in later stages. Given the feature extractor from classification layer $f_D$ and feature weight $\gamma$, we have:
\begin{multline*}
L_{cyc}(G,F,D_X,X,\gamma) = E_{x\sim p_{data}(x)}[\gamma\norm{f_{D_X}(F(G(x)))-f_{D_X}(x)}_1+ \\
(1-\gamma)\norm{F(G(x))-x}_1]
\end{multline*}
Second, a decaying weight $\lambda$ was added to the modified cycle loss, so that the network is stabilizing quickly at early stages, and will not be constrained by cycle loss in later stages. The cycle loss then becomes:
\begin{multline*}
L_{cyc}(G,F,D_X,X,\gamma) = E_{x\sim p_{data}(x)}[D_X(x)[\gamma\norm{f_{D_X}(F(G(x)))-f_{D_X}(x)}_1+ \\
(1-\gamma)\norm{F(G(x))-x}_1]]
\end{multline*}
Last, the modified cycle loss from each direction was further weighted by the output of the corresponding discriminator. It helps with the cases when the generated fake images are unrealistic. The overall feature-weighted loss is:
\begin{multline*}
L(G,F,D_X,D_Y,t) = L_{GAN}(G,D_Y,X,Y) + L_{GAN}(F,D_X,Y,X) + \\
\lambda_t (L_{cyc}(G,F,D_X,X,\gamma_t) + L_{cyc}(F,G,D_Y,Y,\gamma_t) + L_{seg}(S,M,G)
\end{multline*}

\section{Results}
\label{l_result}
The two proposed methods, traditional Gaussian noise injection and CycleGAN, were both evaluated qualitatively, and quantitatively on three volumes of 128 b-scans. 

\subsection{Qualitative Evaluation}
\begin{figure}[H]
    \centering
    \includegraphics[scale=0.35]{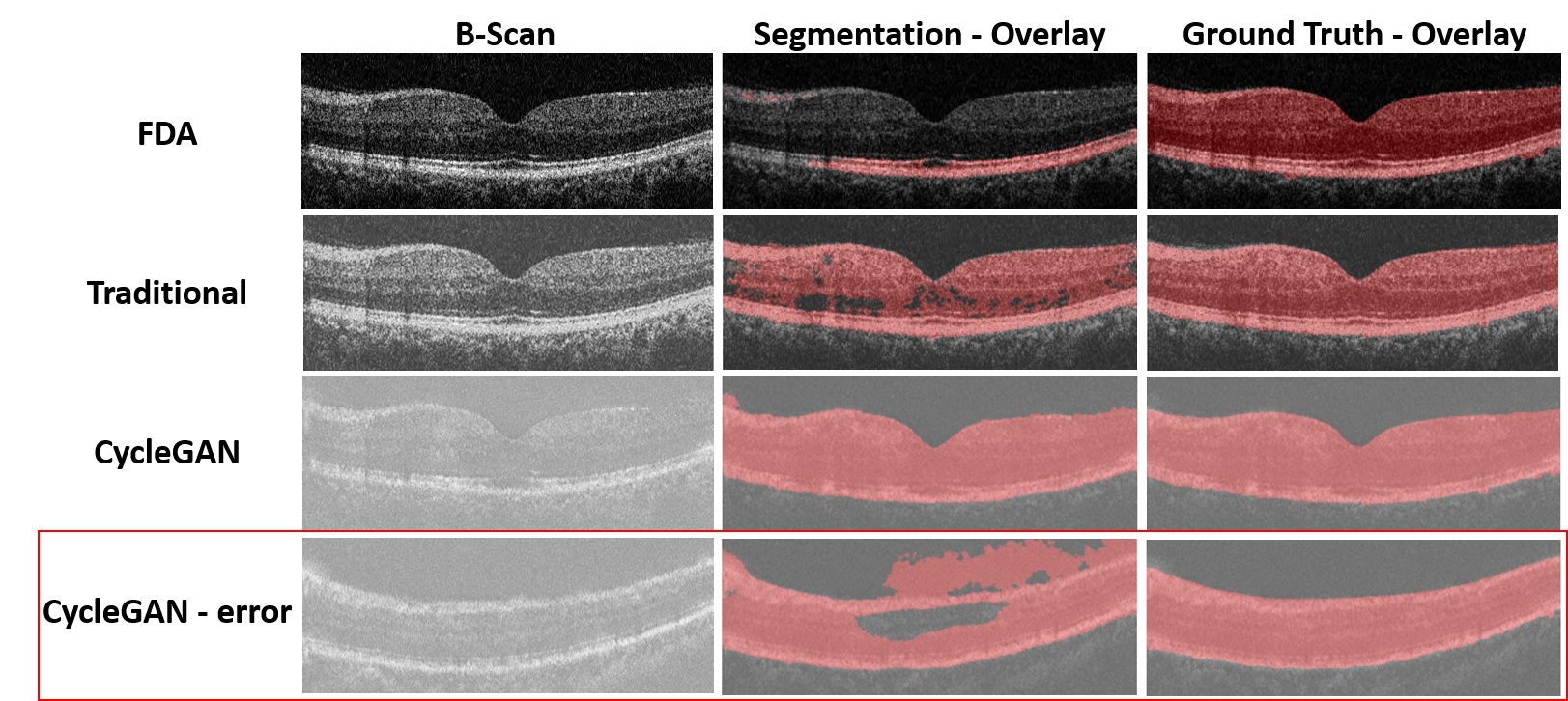}
    \caption{Retinal segmentation (red overlay) of the unprocessed (FDA), Gaussian noise (Traditional), machine learning (\acrshort{cyclegan}) approaches. An example of a poorly segmented B-scan is highlighted by the red box.}
    \label{fig:seg_perf}
\end{figure}

Visually, the segmentation produced from the methods proposed in the previous sections had varying results where the \acrshort{cyclegan} approach outperformed the traditional method, which outperformed the unprocessed FDA B-scans. This trend was consistent throughout all three volumes of the test set, as shown in Figure \ref{fig:seg_perf}. In the left-most column, it can be seen that the intensity of the processed B-scans was overall increased as the method got more complex. The intensity was normalized prior to retina segmentation, thus, the absolute intensity value is not relevant. As shown in Figure \ref{fig:seg_perf}, the segmentation of the unprocessed images is highly inaccurate and only segments the retina near the \acrshort{rpe}. The traditional method of injecting Gaussian noise and intensity shifts performed extremely well in terms of retinal boundaries, but had many false negatives within the segmented retina. Domain adaptation through a \acrshort{cyclegan} resulted in the most visually appealing retina segmentation where the retinal boundaries are fairly consistent and there were rarely retinal pixels classified as background. 

While this trend of visual performance was consistent throughout most of the test volumes, there were outliers where the \acrshort{cyclegan} domain adaptation resulted in poor inference, as shown in the highlighted row of Figure \ref{fig:seg_perf}. While visual inspection suggested a subjective preference for \acrshort{cyclegan} over the traditional method, a quantitative analysis will allow an objective comparison.
\newpage
\subsection{Quantitative Evaluation}
\begin{figure}[H]
    \centering
    \includegraphics[scale=0.5]{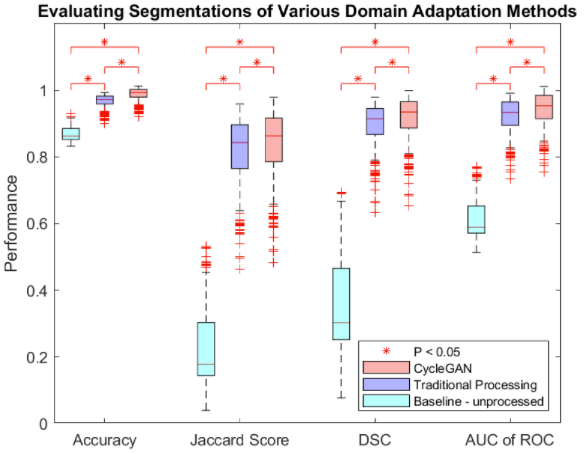}
    \caption{Segmentation evaluation metrics statistical analysis where statistically significant difference (P < 0.05) in means is highlighted by a red asterisk.}
    \label{fig:eval_fig}
\end{figure}

With reference to the Literature, the following performance metrics were calculated between the segmentation and ground truth for quantitative evaluation: accuracy \cite{Metric-all}\cite{Metric-2}, \acrfull{dsc} \cite{Metric-all}\cite{Metric-2}, \acrfull{js} \cite{Metric-all}, and \acrfull{auc} \cite{Metric-all}. The \acrshort{ukb} processing methods resulted in the segmentation performances from tools trained on the RETOUCH dataset, as shown in Table \ref{table:eval_table}. Figure \ref{fig:eval_fig} graphically represents the segmentation performance of each method for all of the above evaluation metrics extensively used in the Literature. 

\begin{table}[H]
    \centering
    \caption{Domain adaptation performance through segmentation evaluation metrics. Formatted as follows: Mean (Standard Deviation).}
    \includegraphics[width=0.6\textwidth,height=2.5cm]{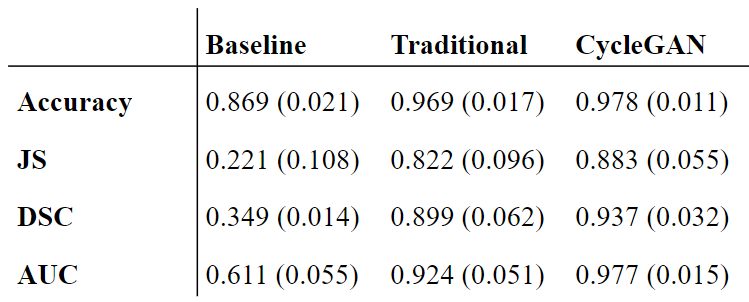}
    \label{table:eval_table}
\end{table}

In the box plot, the light blue, purple, and red distributions represent the unprocessed FDA \acrshort{ukb}, tradition Gaussian noise injection, and \acrshort{ai} \acrshort{cyclegan} approaches, respectively. Statistical significance (P < 0.05) of the difference in means was calculated using two-tailed t-tests and highlighted by a red asterisk. Across all metrics, there was a statistically significant preference from the segmentation tools trained on the RETOUCH data for \acrshort{ukb} adapted through \acrshort{cyclegan} over the traditional method, which in turn had a statistically significant preference over the unprocessed \acrshort{ukb} data. 

\newpage
\section{Conclusion}
\label{l_conclusion}
This study explored the performance of domain adaptation of drastically different data (\acrshort{ukb}) to the domain used to train the \acrshort{oct} retinal layer segmentation tools. The retinal segmentation of the domain adaptation through a \acrshort{cyclegan} was compared to unprocessed, and traditional Gaussian noise injected retinal images. The segmentation on \acrshort{cyclegan}-adapted volumes was shown to be superior in accuracy, \acrshort{dsc}, \acrshort{js}, and \acrshort{auc}. Future steps should be taken to explore the effect of \acrshort{cyclegan} on pathological eyes where the segmentation performance may be reduced. Regardless, the \acrshort{cyclegan} domain-adapted \acrshort{ukb} data resulted in retinal inference with \acrshort{dsc} of 0.937, which is comparable to the Literature which reports \acrshort{dsc} ranging from 0.928 to 0.961 from different retinal layers \cite{LF-UNet}. In this domain adaptation experiment, from the perspective of adapting data to pretrained tools, the \acrshort{cyclegan} was significantly superior to a traditional-intensity-based approach.

\end{document}